\documentstyle[aps,pre,multicol,epsfig,times]{revtex}
\makeatletter
\newlength\abovecaptionskip \newlength\belowcaptionskip
\def\@makecaption#1#2{
 \vskip\abovecaptionskip \sbox\@tempboxa{#1: #2}
 \ifdim \wd\@tempboxa >\hsize #1: #2\par \else \global \@minipagefalse
 \hb@xt@\hsize{\hfil\box\@tempboxa\hfil}
 \fi \vskip\belowcaptionskip} \makeatother

\begin{document}
\title{Power-law persistence and trends in the atmosphere: A detailed study of long temperature records}
\author{ J. F. Eichner,$^{1,2}$ E. Koscielny-Bunde,$^{1,3}$ A. Bunde,$^1$ S. Havlin,$^2$ and
H.-J. Schellnhuber$^{4}$}
\address{$^1$Institut f\"ur Theoretische Physik III, Universit\"at Giessen, D-35392 Giessen, Germany}
\address{$^2$Minerva Center and Department of Physics, Bar Ilan University, Ramat-Gan, Israel}
\address{$^3$Potsdam Institute for Climate Research, D-14412 Potsdam, Germany}
\address{$^4$Tyndall Centre for Climate Change Research, University of East Anglia, Norwich NR4 7TJ,
United Kingdom}
\date{submitted: 12 December 2002}
\draft\maketitle\begin{multicols}{2}[%
\begin{abstract}
We use several variants of the detrended fluctuation analysis to study the appearance of
long-term  persistence in temperature records, obtained at 95 stations all over the globe. Our
results basically confirm earlier studies. We find that the persistence, characterized by the
correlation $C(s)$ of temperature variations separated by $s$ days,  decays for large $s$ as a
power law, $C(s)\sim s^{-\gamma}$. For continental stations, including  stations along the 
coastlines, we find that $\gamma$ is always close to 0.7. For stations on islands, we find that
$\gamma$ ranges between 0.3 and 0.7, with a maximum at $\gamma=0.4$. This is consistent with
earlier studies of the persistence in sea surface temperature records where $\gamma$ is close to
0.4. In all cases, the exponent $\gamma$ does not depend on the distance of the stations to the 
continental coastlines. By varying the degree of detrending in the fluctuation analysis we obtain 
also information about trends in the temperature records. 
\end{abstract}
]
	 
\section{Introduction}
The persistence of weather states on short terms is a well-known phenomenon: A warm day is more
likely to be followed by a warm day than by a cold day and vice versa. The trivial forecast, that
the weather of tomorrow is the same as the weather  of today, was in previous times often used as
a ``minimum skill'' forecast for assessing the usefulness of short-term weather forecasts. The
typical time scale for weather changes is  about 1 week, a time period that corresponds to the
average duration of so-called ``general weather regimes" or ``Grosswetterlagen'', so this type of
short-term persistence usually stops after about 1 week. On larger scales, other types of
persistence occur. One of them is related to circulation patterns associated with blocking
\cite{B8}.  A blocking situation occurs when a very stable high pressure system is established
over a particular region and remains in place for several weeks. As a result the weather in the
region of the high remains fairly persistent throughout this period. It has been argued recently 
\cite{SCAF} that this short-term persistence regime may be linked to solar flare intermittency.
Furthermore, transient low
pressure systems are deflected around the blocking high so that the region downstream of the high
experiences a larger than usual number of storms. On even longer terms, a source for weather 
persistence might be slowly varying external (boundary) forcing such as sea surface temperatures
and anomaly patterns. On the scale of months to seasons, one of the most
pronounced phenomena is the El Nino southern oscillation event which occurs every 3--5 years 
and which strongly affects  the weather over the tropical Pacific as well as over North America
\cite{SienceOfDisasters}.

The question is, {\it how} the persistence that might be generated by very different mechanisms 
on different time scales decays with time $s$. The answer to this question is not easy. 
Correlations, and in particular long-term correlations, can be masked by trends that are
generated, e.g., by the well-known urban warming. Even uncorrelated data in the presence of 
long-term trends may look like correlated ones, and, on the other hand, long-term correlated data
may look like uncorrelated data influenced by a trend.

Therefore, in order to distinguish between trends and correlations one needs methods that can
systematically eliminate trends. Those methods are available now: both wavelet techniques (WT)
(see, e.g., Refs. \cite{WT1,WT2,ARNEODO1,ARNEODO2})  and detrended fluctuation  analysis (DFA) (see,
e.g., Refs. \cite{DNA1,HEART3,KANT,HU}) can systematically eliminate trends in the data and thus
reveal intrinsic dynamical properties such as distributions, scaling and long-range  correlations
very often masked by nonstationarities. 

In a previous study \cite{EVA}, we have used DFA and WT to study temperature   correlations in
different climatic zones on the globe. The analysis focused on 14 continental stations, several
of them were located along coastlines. The results  indicated that the temperature variations
are long-range power-law correlated above some  crossover time that is of the order of 10 days.
Above the crossover time, the persistence, characterized by the autocorrelation $C(s)$ of
temperature variations separated by $s$ days, decayed as  
$$C(s)\sim s^{-\gamma}, \eqno(1)$$
where, most interestingly, the exponent $\gamma$ had roughly the same value $\gamma\cong 0.7$ for
all continental records. Equation (1) can be used as a test bed for global climate models
\cite{GOVINDAN}.

More recently, DFA was applied to study temperature   correlations in the sea surface
temperatures \cite{Roberto}. It was found that the temperature  autocorrelation function $C(s)$
again decayed by a power law, but with an exponent $\gamma$ close  to 0.4, pointing towards a
stronger persistence in the oceans than in the continents.

In this paper, we considerably extend our previous analysis to study systematically temperature
records of 95 stations. Most of them are on the continents, and several of them are on islands.
Our results are actually in line with both earlier papers and in agreement with conclusions drawn
from independent type of analysis by  several groups \cite{EVA2,P1,T1}. We find that the
continental records, including those on coastlines, show power-law persistence with $\gamma$
close to 0.7, while the island records show power-law correlations with $\gamma$ around 0.4.  By
comparing different orders of DFA that differ in the way trends are eliminated, we could also
study the presence of trends in the records that lead to a warming of the atmosphere. We find
that pronounced trends occur mainly at big cities and can be probably attributed to urban
growth. Trends that cannot be attributed to urban growth occur in half of the island stations
considered and on summit stations in the Alps. A majority of the stations showed no
indications of trends.

The article is organized as follows. In Sec. II, we describe the detrending analysis used in
this paper, the DFA. In Sec. III, we present the result of this
analysis. Sec. IV concludes the paper with a discussion.

\section{The methods of analysis}

Consider a record $T_i$, where the index $i$ counts the days in the record, $i=1,2,...,N$. The
$T_i$ represent the maximum daily temperature, measured  at a certain meteorological station. For
eliminating the periodic seasonal trends, we concentrate on the  departures of $T_i$, $\Delta
T_i=T_i - \overline T_i$, from their mean daily value $\overline T_i$ for each calendar date $i$,
say, 2nd of March, which has been obtained by averaging  over all years in the record.

Quantitatively, correlations between two $\Delta T_i$ values separated by $n$ days are defined by
the (auto) correlation function 
$$C(n) \equiv \langle \Delta T_i \Delta T_{i+n}\rangle={1\over
N-n} \sum_{i=1}^{N-n} \Delta T_i\Delta T_{i+n}.   \eqno(2)$$ 
If $\Delta T_i$ are uncorrelated, $C(n)$ is zero for $n$ positive. If correlations exist 
up to a certain number of days $n_\times$, the correlation function will be positive up to 
$n_\times$ and vanish above $n_\times$. A direct calculation of $C(n)$ is hindered by the level 
of noise present in the finite records, and by possible nonstationarities in the data.

To reduce the noise we do not calculate $C(n)$ directly, but instead study the ``profile''
$$Y_m=\sum_{i=1}^m \Delta T_i. $$ 
We can consider the profile $Y_m$ as the position of a random
walker on a linear chain after $m$ steps. The random walker starts at the origin and performs, in
the $i$th step, a jump of length $\Delta T_i$ to the right, if $\Delta T_i$ is positive, and to
the left, if $\Delta T_i$ is negative. The fluctuations ${F^2(s)}$ of the profile, in a given
time window of size $s$, are related to the correlation function $C(s)$. For the relevant case
(1) of long-range power-law correlations, $C(s)\sim s^{-\gamma},\quad 0<\gamma<1,$ the
mean-square fluctuations $\overline{F^2(s)}$, obtained  by averaging over many time windows of
size $s$ (see below) asymptotically increase by a  power law \cite{BARAB}:
$$
\overline{F^2(s)}\sim s^{2\alpha}, \quad\alpha=1-\gamma/2.\eqno(3)
$$
For uncorrelated data (as well as for correlations decaying faster than $1/s$), we have
$\alpha=1/2$.

For the analysis of the fluctuations, we employ a hierarchy of methods that differ in the way the
fluctuations are measured and possible trends are eliminated (for a detailed description of the
methods we refer to Ref. \cite{KANT}).

(i) In the simplest type of fluctuation analysis (DFA0) (where trends are not going to be
eliminated), we determine in each window the mean value of the profile. The variance of the
profile from this constant value represents the square of the fluctuations in each window.

(ii) In the {\it first order} detrended fluctuation analysis (DFA1), we determine in each window
the best linear fit of the profile. The variance of the profile from this straight line
represents the square of the fluctuations in each window.

(iii) In general, in the $n$th order DFA (DFAn) we determine in each window the best $n$th
order polynomial fit of the profile. The variance of the profile from these best $n$th-order
polynomials represents the square of the fluctuations in each window.

By definition, DFA0 does not eliminate trends, while DFAn eliminates trends of order $n$ in the
profile and $n-1$ in the original time series. Thus, from the comparison of fluctuation functions
$F(s)$ obtained from different methods one can learn about both, long-term correlations and the
influence of trends. 

The DFA method is analogous to wavelet techniques that also eliminate polynomial
trends systematically. For a detailed review of the method, see Refs. \cite{ARNEODO1,ARNEODO2}. The 
conventional techniques such as the direct evaluation of $C(n)$, the rescaled range analysis (R/S) 
introduced by Hurst (for a review, see, e.g., Ref. \cite{FEDER}) or the power spectrum method 
\cite{P1,T1,TURCOTTE,LOVEJOY} can only be applied on stationary records. In the presence 
of trends they may overestimate the long-term persistence exponent. The R/S method is somewhat similar 
to the DFA0 analysis.
 
\section{Analysis of temperature records}

\noindent Figure 1 shows the results of the DFA analysis  of the daily temperatures (maximum or
mean values) $T_i$ of the following weather stations (the length of the records is written within
the parentheses): (a) Vienna (A, 125 yr), (b) Perm (RUS, 113 yr), (c) Charleston (USA, 127 yr), and
(d) Pusan (KOR, 91 yr). Vienna and Perm have continental climate, while Charleston and Pusan are
close to coastlines.

In the log-log plots the DFA1--3 curves are (except at small $s$ values) approximately straight
lines. For both the stations inside the continents and along coastlines the slope is
$\alpha\cong 0.65$.  There exists a natural crossover (above the DFA crossovers at very small
times) that can be best estimated from DFA0 \cite{Crossover}. As can be verified easily, the 
crossover occurs roughly at $s_\times=10$ days, which is the order of magnitude  for a typical 
Grosswetterlage. Above $s_\times$, there exists long-range persistence expressed by the  power-law 
decay of the correlation function with an exponent $\gamma = 2 -2 \alpha \cong 0.7$. 

\begin{figure}
\noindent\includegraphics[width=8.5cm]{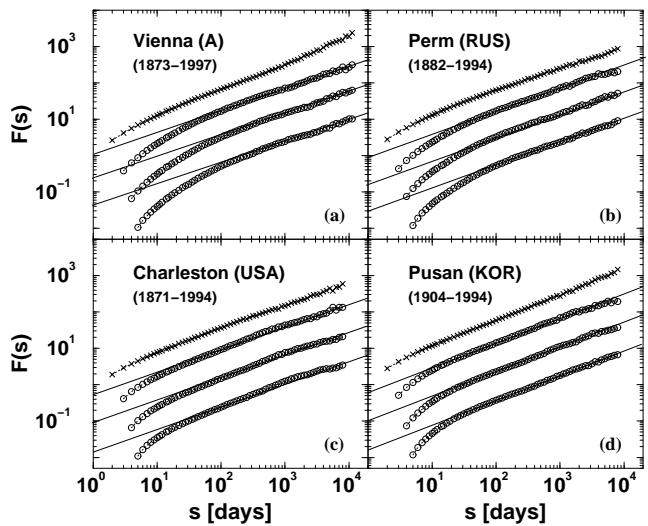}
\parbox{8.5cm}{\caption[]{\small
Analysis of daily temperature records of four representative weather stations on continents. The four
figures show the fluctuation functions obtained by DFA0, DFA1, DFA2, and DFA3 (from top to 
bottom) for the four sets of data. The slopes are $0.64 \pm 0.02$ (Vienna), $0.62 \pm 0.02$ (Perm), 
$0.63 \pm 0.02$ (Charleston), and $0.67 \pm 0.02$ (Pusan). Lines with these slopes are plotted in the
figures. The scale of the fluctuation functions is arbitrary.}} 
\end{figure}

\begin{figure}
\noindent\includegraphics[width=8.5cm]{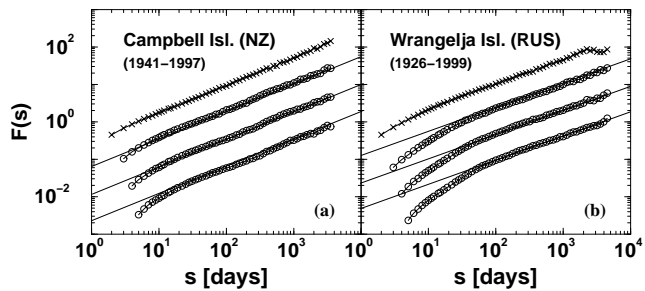}
\parbox{8.5cm}{\caption[]{\small
Analysis of daily temperature records of two representative weather
stations on islands. The DFA curves are arranged as in Fig. 1. The slopes are $0.71 \pm 0.02$ 
(Campbell) and $0.65 \pm 0.02$ (Wrangelja). Lines with these slopes are plotted in the figures.}}
\end{figure}

Figure 2 shows the results of the DFA analysis of the  daily temperatures for two island
stations: Wrangelja and Campbell Islands. Wrangelja Island is a large island between the East
Siberian Sea and the Chukchi Sea. During the winter season, large parts of the water  surrounding
the island are usually frozen. Campbell Island is a small island belonging to New Zealand. Again,
in the double logarithmic presentation, all DFA1--3 fluctuation functions are straight lines, but
the slopes differ. While for Wrangelja the slope is 0.65, similar to the land stations shown
before,  the slope for Campbell Island is  significantly larger, close to 0.8 (corresponding to
$\gamma=0.4$).

It can be seen from Figs. 1 and 2 that sometimes the DFA0 curves have a larger slope than the
DFA1--3 curves, and that usually the curves of DFA2 and DFA3 have the same slope for large $s$ values.
The fact that the DFA0 curve has a higher exponent indicates the existence of trends by which
the long-term correlations are masked. Calculations using DFA0 alone will yield a  higher
correlation exponent  and thus lead to a spurious overestimation of the long-term persistence.
The fact that the DFA2 and DFA3 curves show the same asymptotic behavior indicates that possible
nonlinearities in the trends are not significant. Otherwise the DFA2 curve  (where only linear
trends are eliminated) would show an asymptotic behavior different from DFA3.

\begin{figure}
\noindent\includegraphics[width=8.5cm]{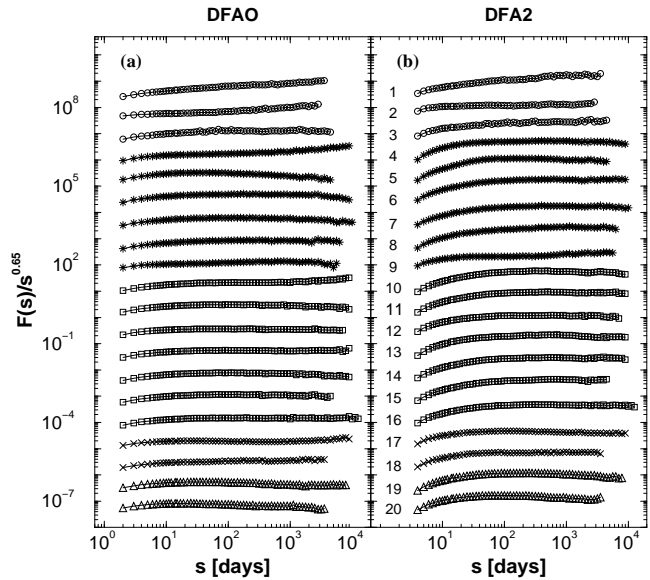}
\parbox{8.5cm}{\caption[]{\small
Fluctuation analysis by DFA0 and DFA2 of daily temperature records of 20 representative weather
stations: (1) Thursday Island (AUS, 53 yr), (2) Koror Island (USA, 54 yr), (3) Raoul Island (USA, 54
yr), (4) Hong Kong (C, 111 yr), (5) Anadir (RUS, 101 yr), (6) Hamburg (D, 107 yr), (7) Plymouth (GB, 122
yr), (8) Feodosija (UA, 113 yr), (9) Wellington (NZ, 67 yr), (10) Jena (D, 175 yr), (11) Brno (CZ, 128
yr), (12) Chita (RUS, 114 yr),  (13) Tashkent (USB, 119 yr), (14) Potsdam (D, 115 yr), (15) Minsk (WY,
113 yr), (16) Oxford (GB, 155 yr), (17) Cheyenne (USA, 123 yr), (18) Kunming (C, 49 yr), (19)
Wuxqiaoling (C, 40 yr), and (20) Zugspitze (D, 98 yr). Stations 1--3 are on islands, stations 4--9 
are on coastlines, and stations 10--20 are inland stations, among them two stations (19 and 20) are 
on summits. The scales are arbitrary. To reveal that the exponents $\alpha$ are close to 0.65, we have 
divided the fluctuation functions by $s^{0.65}$.}}
\end{figure}

By comparing the DFA0 curves with the DFA2 curves, we can learn more about possible trends.
Usually the effect of trends is seen as a crossover in the DFA0 curve. Below the crossover, the
slopes of DFA0 and DFA2 are roughly the same, while above the crossover the DFA0 curve bends up.
Large trends are characterized by a short crossovertime $s_\times$ and a large difference in  the
slopes between DFA0 and DFA2 (for a general discussion see Refs. \cite{KANT} and \cite{HU}).  A  nice
example for this represents Vienna, where the DFA0 curve shows a pronounced crossover at about
3 yr.  Above this crossover, the DFA0 curve bends up considerably, with an effective slope close to
0.8. For Pusan, the trend is less pronounced, and for Perm, Charleston, and  the two islands we do
not see indications of  trends.

To reveal the presence of long-term correlations and to point out possible trends, we have
plotted in Fig. 3(a) the DFA0 curves and in Fig. 3(b) the DFA2 curves for 20 representative
stations around the globe. For convenience, the fluctuation functions have been divided by
$s^{0.65}$. We do not show results for those stations that were analyzed in Ref. \cite{EVA}. Figure
3(b) shows again that continental and coastline stations have roughly the same fluctuation
exponent $\alpha\cong 0.65$, while islands may also have higher exponents. It seems that stations
at peaks of high mountains [here we show Zugspitze (D, 98 yr, no. 19) and Wuxqiaoling (C, 40 yr,
no. 20)] have a slightly lower exponent. 

From the 26 stations shown in Figs. 1--3, 8 show a larger exponent in the DFA0 treatment than in
the DFA2 treatment. These stations are Thursday Island (no. 1 in Fig. 3), Koror Island (no. 2 in
Fig. 3), as well as Vienna [Fig. 1(a)], Pusan [Fig. 1(d)],  Hong Kong  (no. 4 in Fig. 3), Jena (no.
10 in Fig. 3), Cheyenne (no. 17 in Fig. 3), and Zugspitze (no. 19 in Fig. 3).  The other 18
stations do not show a difference in the exponents for DFA0 and DFA2,  which suggests that the
trends are either zero or too small to be detected by this sensitive method. We observe the
largest trends for Hong Kong, Vienna, and Jena, where in all cases the crossover in the DFA0
curve is around 3 yr  and the final slope is between 0.75 and 0.8. It is obvious that the greatest
part of this warming is due to  the urban growth of theses cities. Regarding the two islands,
Koror shows a pronounced trend with a crossover time below 1 yr, while the trend we observe  for
Thursday Island is comparatively weak. It is not likely that the trends on the islands can be
attributed to urban warming. 

\begin{figure}
\noindent\includegraphics[width=8.5cm]{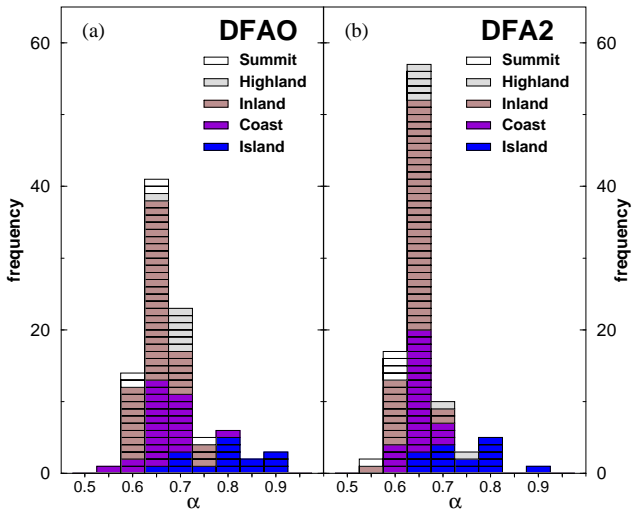}
\parbox{8.5cm}{\caption[]{\small
Histogram of the values of the fluctuation exponents $\alpha$ obtained (a) from DFA0 where trends
are not eliminated and (b) from DFA2 where linear trends are eliminated systematically on all
time scales.}}
\end{figure}

Figure 4 summarizes our results for all the stations analyzed. Fig. 4(a) shows the histogram for the
values of the exponent $\alpha$ obtained by DFA0, while Fig. 4(b) shows the corresponding
histogram obtained by DFA2. Both histograms are quite similar. For DFA2 the average exponent
$\alpha$ is $0.66 \pm 0.06$ and for DFA0 it is  $0.68 \pm 0.07$.  The maxima become sharper when
the islands are eliminated from the figures. The slight shift towards larger $\alpha$ values in
DFA0 is due to trends. The magnitude of the trends can be roughly characterized by the difference
$\delta\alpha$ of the slopes of DFA0 and DFA2. We found that 7 of the 15 island stations and 54
of the 80 continental stations showed no significant trend, with  $\delta\alpha \le 0.02$. We
observed a small trend, with $0.03 \le \delta\alpha \le 0.05$, for 3 island and 9 continental
stations. A pronounced trend, with $\delta\alpha \ge 0.06$, was found for 5 island and 13
continental stations. Among these 13 stations are Hong Kong, Bordeaux, Prague, Seoul, Sydney,
Urumchi, Swerdlowsk, and Vienna, where a large part of the warming can be attributed to the urban
growth of the cities in the last century. Two of these stations [S\"antis (CH) and Sonnblick (A)]
are on top of high mountains.

\begin{figure}
\noindent\includegraphics[width=8.5cm]{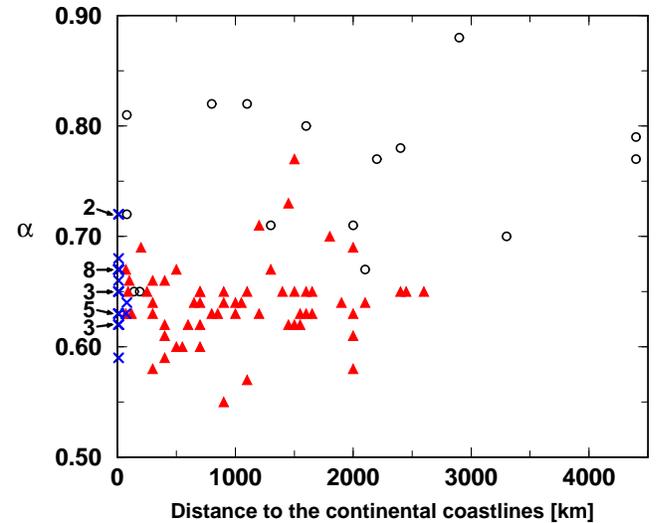}
\parbox{8.5cm}{\caption[]{\small
The scaling exponent $\alpha$ as a function of the distance $d$ between the stations and the 
continental coastlines, for island stations ($\circ$), continental stations ($\triangle$), and 
coastline stations ($\times$). Many of the coastline stations ($d=0$) have the same $\alpha$ value, 
and we indicated their number in the figure.}}
\end{figure}

Since the island stations have exponent $\alpha$ larger than the continental stations, it is 
likely that the long-term persistence originates by the coupling of the atmosphere to the oceans. 
Thus one might expect that for island stations $\alpha$ will increase with the distance to the 
continents, and for continental stations $\alpha$ will decrease with the distance to the 
coastline. To test if the exponent $\alpha$ depends on the distance $d$ to the continental 
coastlines, we plotted in Fig. 5 $\alpha$ as a function of $d$ for both islands and continental 
stations. It is remarkable that islands far away from the continents do not show a larger 
exponent than islands close to the coastlines, and inner-continental stations far from the ocean 
do not show smaller exponents than coastline stations. This second result is in disagreement with 
a recent claim that $\alpha = 0.5$ for inner-continental stations far away from the oceans 
\cite{FB-PRL}.

\section{Discussion}
In this paper, we have used a hierarchy of detrending analysis methods (DFA0--DFA3) to study long
temperature records around the globe. We concentrated mainly on those areas on the globe (North
America, Europe, Asia and Australia) where long records are available. The main results of the
study are the following.

(i) The temperature persistence decays, after a crossover time that is typically of the order of
the duration of a Grosswetterlage, by a power law, with an exponent $\alpha$ that has a very
narrow distribution for continental stations. The mean value of the exponent is close to  0.65,
in agreement with earlier calculations based on different methods \cite{EVA,EVA2,P1,T1}.

(ii) On islands, the exponent shows a broader distribution, varying from 0.65 to 0.85, with an
average value close to 0.8. This finding is in qualitative agreement with the results of a recent
analysis of sea surface temperature records, where also long-term persistence  with an average
exponent close to 0.8 has been found \cite{Roberto}. Since the oceans cover more than 2/3 of the
globe, one may expect that also the mean global temperature is characterized by long-term
persistence, with an exponent close to 0.8.

(iii) In the vast majority of stations we did not see indications for a global warming of the
atmosphere.  Exceptions are mountain stations in the Alps [Zugspitze (D), S\"antis (CH), and
Sonnblick (A)], where urban warming can be excluded. Also, in half of the islands we studied, we
found pronounced trends that most probably cannot be attributed to urban warming. Most of the
continental stations where we observed significant trends are large cities where probably the
fast urban growth in the last century gave rise to temperature increases. 

When analyzing warming phenomena in the atmosphere, it is essential to employ methods that can 
distinguish, in a systematic way, between trends and long-term correlations in contradistinction 
to a number of conventional schemes that have been applied in the past. These schemes run the risk 
of mixing up the correlatedness of natural climate system variability with entire regime shifts 
enforced by anthropogenic interference through greenhouse gas emissions. The fact that we found it 
difficult to discern warming trends at many stations that are not located in rapidly developing 
urban areas may indicate that the actual increase in global temperature caused by anthropogenic
perturbation is less pronounced than estimated in the last IPCC (Intergovernmental Panel for 
Climate Change) report \cite{IPCC}.

\section*{Acknowledgments}
We are grateful to  Professor S. Brenner for very useful discussions. We would like to acknowledge
financial support by the Deutsche Forschungsgemeinschaft and the Israel Science
Foundation.

\end{multicols}
\end{document}